\documentclass[letter,traditabstract]{aa} 
\usepackage{graphicx}
\usepackage[varg]{txfonts}
\usepackage{natbib}
\bibpunct{(}{)}{;}{a}{}{,} 

\begin{document}

   \title{Solution to the problem of the surface gravity distribution of cool DA white dwarfs from improved 3D model atmospheres}

   \author{P.-E. Tremblay\inst{1}
          \and
          H.-G. Ludwig\inst{2}
          \and
          M. Steffen\inst{3}
          \and
          P. Bergeron\inst{1}
          \and
          B. Freytag\inst{4}
          }

   \institute{D\'epartement de Physique, Universit\'e de Montr\'eal, C.P.~6128, 
Succ.~Centre-Ville, Montr\'eal, QC H3C 3J7, Canada\\
              \email{tremblay@astro.umontreal.ca, bergeron@astro.umontreal.ca}
         \and
            Zentrum f\"ur Astronomie der Universit\"at Heidelberg, Landessternwarte, 
            K\"onigstuhl 12, 69117 Heidelberg\\
             \email{hludwig@lsw.uni-heidelberg.de}
         \and
            Leibniz-Institut f\"ur Astrophysik Potsdam, An der Sternwarte 16, D-14482 Potsdam, Germany\\
             \email{msteffen@aip.de}
         \and
           CNRS, Universit\'e de Lyon, \'Ecole Normale Sup\'erieure de Lyon, 46 all\'ee d'Italie, F-69364 Lyon Cedex 7\\
             \email{bernd.freytag@ens-lyon.fr}
             }

   \date{Received ..; accepted ..}
 
  \abstract {The surface gravities of cool ($T_{\rm eff} < 13,000$ K)
  hydrogen-atmosphere DA white dwarfs, determined from spectroscopic
  analyses, are found to be significantly higher than the canonical
  value of $\log g \sim 8$ expected for these stars. It was
  recently concluded that a problem with the treatment of convective
  energy transport within the framework of the mixing-length theory
  was the most plausible explanation for this high-$\log g$
  problem. We pursue the investigation of this discrepancy by
  computing model spectra of cool convective white dwarfs from a small
  sequence (11,300 K $< T_{\rm eff} <$ 12,800 K) of 3D hydrodynamical
  model atmospheres, which feature a sophisticated treatment of
  convection and radiative transfer. Our approach is to proceed with a
  differential analysis between 3D and standard 1D models. We find
  that the 3D spectra predict significantly lower surface gravities,
  with corrections of the right amplitude as a function of effective
  temperature to obtain values of $\log g \sim 8$ on average. We
  conclude that the surface gravity distribution of cool convective DA
  white dwarfs is much closer to that of hotter radiative objects when
  using, for the treatment of the convection, 3D models instead of the
  mixing-length framework.}{}{}{}{}


   \titlerunning{Solution to the problem of the DA white dwarf $\log g$ distribution from 3D model atmospheres}
   \authorrunning{Tremblay et al.}
   \maketitle

\section{Introduction}

\begin{figure}[!h]
\begin{center}
\includegraphics[bb=18 290 592 602,width=3.8in]{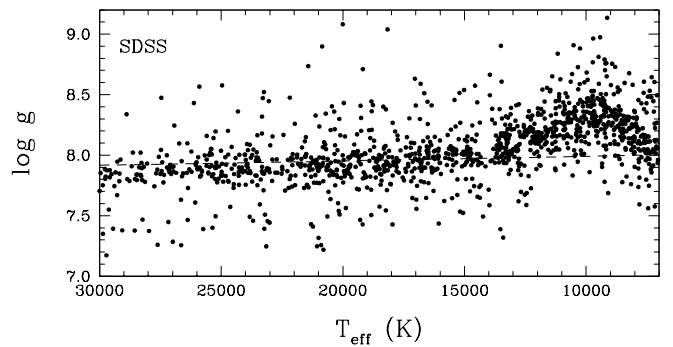}
\caption{Surface gravity distribution as a function of $T_{\rm eff}$ for the SDSS
sample; see \citet{TB11} for details on
these determinations. An evolutionary model from \citet{fontaine01} at
the median mass of the sample (0.59 $M_{\sun}$) is shown as a dashed line.
\label{fg:f1}}
\end{center}
\end{figure}

The shape of Balmer lines in hydrogen-atmosphere DA white dwarfs is
particularly sensitive to variations of the atmospheric
parameters ($T_{\rm eff}$ and $\log g$). Therefore, the spectroscopic
technique, which consists in comparing the observed line profiles of
the Balmer series with the predictions of detailed model atmospheres,
is by far the most accurate method for determining the atmospheric
parameters. This technique was initially applied to a sample of 37
cool ($T_{\rm eff} < 13,000$ K) DA white dwarfs by \citet{bergeron90}
who showed that the $\log g$ values measured for these white dwarfs
were significantly higher than the expected canonical value of $\log g
\sim 8$, which is about the mean value determined later for hotter DA
stars \citep[e.g.,][]{bergeron92}. This discrepancy is now observed in
all large spectroscopic surveys of DA white dwarfs
\citep[e.g.,][]{SDSS,gianninas} and no satisfactory explanation has
been reported until now. In Fig.~\ref{fg:f1}, we present our current
view of this high-$\log g$ problem for the spectroscopically
identified DA stars in the Sloan Digital Sky Survey (SDSS). This
long-standing problem implies that we cannot obtain reliable
atmospheric parameters determined from spectroscopy at the cool end of
the white dwarf evolutionary sequence. Since the problem manifests
itself for white dwarfs older than $\sim$1 Gyr, it has important
implications on our ability to use these objects as cosmochronometers
and distance indicators \citep{winget87,fontaine01}.

Since white dwarf stars are expected to cool at constant mass and
almost constant radius, the sudden and significant increase in $\log
g$ cannot be explained in terms of simple astrophysical
arguments. \citet[][hereafter TB10; see also
\citealt{koester09}]{TB10} made an extensive review of the possible
solutions, which range from inadequate assumptions about the
composition of these stars to inaccuracies in the model atmosphere
calculations. The non-detection of He~\textsc{i} lines in
high-resolution Keck observations of cool DA stars by TB10 ruled out
the systematic presence of helium in the atmospheres, which would
mimic higher spectroscopic $\log g$ determinations, as initially
proposed by \citet{bergeron90}. TB10 concluded -- as \citet{koester09}
did -- that a problem with the treatment of convective energy
transport, currently based on the mixing-length theory (MLT;
\citealt{MLT}), is the only viable explanation for the high-$\log g$
problem.

The MLT approximation is a phenomenological description of the complex
convective fluid movements in a star. According to this model, {\it
bubbles} transport energy in the convection zone over a characteristic
distance -- the mixing length -- where they dissolve and release their
extra energy in the environment. Even though the MLT framework,
properly calibrated, provides a reasonable description of white dwarf
model atmospheres (TB10 and references therein), even small changes in
the convective efficiency, or equivalently the mixing-length
parameter, can have a significant impact on the predicted line
profiles \citep[see, e.g., Fig.~3 of][]{bergeron95} in the regime
where the high-$\log g$ problem is most important. This comes from the
fact that convective energy transport may affect dramatically the
temperature and pressure structures of the atmosphere, and
consequently the shape of the broad hydrogen lines, formed at
different depths in the atmosphere. The observations reveal that the
high-$\log g$ problem seems to be related to the line profiles only,
and not to the continuum flux (as demonstrated from photometric
analyses of these stars; see, e.g., TB10 and \citealt{koester09}), and
that the problem occurs when convection becomes important in the
photosphere (TB10). This is consistent with the MLT framework being
the source of the problem.

A more detailed multidimensional radiation-hydrodynamics (RHD)
treatment of convective motions in DA white dwarfs, based on first
physical principles, has already been studied by \citet{ludwig94},
\citet{steffen95}, and \citet{freytag96}. These studies revealed that
the differences between 2D spectra and those computed from standard 1D
hydrostatic models were qualitatively small. However, a visual
inspection of Fig.~13 from TB10 reveals that the model predictions do
not need to change by much to rectify the high-$\log g$ problem. These
2D calculations have not been pursued further, and in the meantime,
full three dimensional RHD simulations (hereafter 3D models) have
become more sophisticated, and accurate enough to make realistic
predictions (see, e.g., \citealt{asplund00} and references therein).

In this work, we present an updated investigation of the
differences between standard DA spectra and those calculated from 3D
model atmospheres. In Sect. 2, we describe our computation of a
small number of 3D model atmospheres for cool convective DA white
dwarfs. We then address, in Sect. 3, the status of the high-$\log g$
problem from a purely differential analysis between the Balmer line
profiles predicted by the 3D models, and those computed from 1D models
with the same underlying microphysics.

\section{Improved 3D model atmospheres}

We computed a small set (11,300 K $< T_{\rm eff} <$ 12,800 K) of pure-hydrogen
3D model atmospheres at $\log g = 8$ with the CO$^{5}$BOLD
code\footnote{www.astro.uu.se/$\sim$bf/co5bold$\_$main.html}
\citep{freytag02,wedemeyer04}. This code solves the time-dependent
hydrodynamical equations (conservation of mass, momentum, and energy)
for a fully compressible fluid with a finite-volume approach, coupled
to the radiative transfer. The DA white dwarfs investigated here have
a shallow convective zone, which can be included entirely in the
simulation domain (for the vertical direction). This allows for
dynamically stable layers at the top and bottom of the domain, and
accounts for the entire overshooting layer. The lateral boundaries are
periodic, and the radiative flux entering the domain at the bottom
defines the effective temperature. We rely on a standard stellar
equation of state, and the opacities are the same as those used in the
simulations of \citet{ludwig94}. We computed both gray and non-gray
models, the latter using 7 opacity bins for the radiative transfer
\citep{ludwig94}. Table~\ref{tab1} presents the basic properties for
our set of models. In Fig.~\ref{fg:f2}, we illustrate the structure of the
non-gray 11,975 K model. We find that our new models are qualitatively
similar to those computed by \citet{ludwig94} and
\citet{steffen95} in terms of the properties of the convective zone
(i.e., depth and maximum velocities).

 \begin{table}[!h]
 \caption{Parameters of our 3D model atmospheres}
 \label{tab1}
 \begin{center}
 \begin{tabular}{lcccc}
\hline
\hline
$T_{\rm eff}$ & $\log g$ & Opacity bins & Time\tablefootmark{a} & $\vert dF \vert$\tablefootmark{b}\\
(K) &  &  & (s) & ($\%$) \\
\hline

\hline
\noalign{\smallskip}
\multicolumn{5}{c}{ Non-gray models} \\
\noalign{\smallskip}
\hline
11,300 & 8.0 & 7 & 10 & 2.46 \\
11,975 & 8.0 & 7 & 10 & 0.07 \\
12,390 & 8.0 & 7 & 10 & 0.16 \\
12,800 & 8.0 & 7 & 10 & 0.02 \\
\hline
\noalign{\smallskip}
\multicolumn{5}{c}{ Gray models} \\
\noalign{\smallskip}
\hline
11,350 & 8.0 & 1 & 40 & 1.20 \\
11,975 & 8.0 & 1 & 10 & 0.70 \\
12,390 & 8.0 & 1 & 15 & 0.03 \\
12,800 & 8.0 & 1 & 15 & 0.06 \\
\hline

\end{tabular} 
\end{center} 
\tablefoottext{a}{Total stellar time.}
\tablefoottext{b}{
Maximum flux error, defined as (${\rm max}\vert
F(z)/F^*-1\vert$), taken over the whole depth of
the computation box. $F^*$ is the nominal flux 
$\sigma T_{\rm eff}^4$, and $F(z)$ is the total
energy flux (radiative, convective, kinetic, viscous
and potential fluxes) averaged over 2 seconds
(or 5 seconds for the coolest gray and non-gray models) at the
end of the simulation. Note that the maximum flux 
error generally occurs at the
base of the simulation and that in the line forming layers, 
the flux
error is much smaller.}

\tablefoot{The box size is 7.45 km (150
grid points) for both horizontal directions and 4.17 km for the
vertical direction (100 grid points).}  
\end{table}

\onlfig{2}{
\begin{figure*}[!]
\begin{center}
\vspace{-2cm}
\includegraphics[width=4.2in,angle=90]{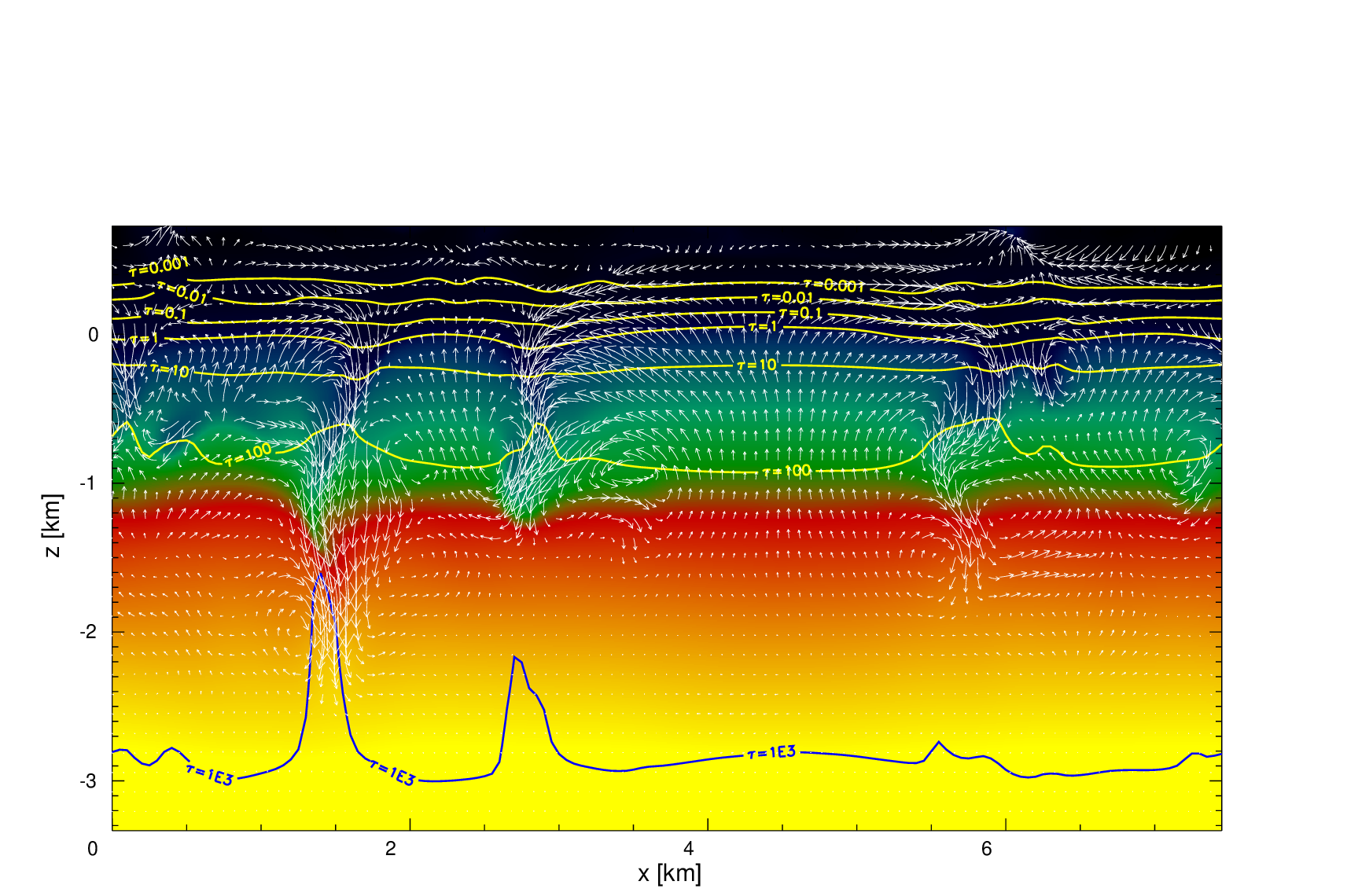}
\caption
{Temperature structure for the 3D white dwarf model at $T_{\rm eff}=11,975$~K
and $\log g = 8$, for a slice in the horizontal-vertical $xz$ plane
through a box with coordinates $x,y,z$ (in km). The temperature is
color coded from 60,000~K (yellow) to 8000~K (blue). The arrows
represent relative convective velocities (integration time of 20 ms),
while thick lines correspond to contours of constant Rosseland optical
depth.
\label{fg:f2}}
\end{center}
\end{figure*}
}

\begin{figure}[!]
\begin{center}
\includegraphics[bb=58 183 672 648, width=4.5in]{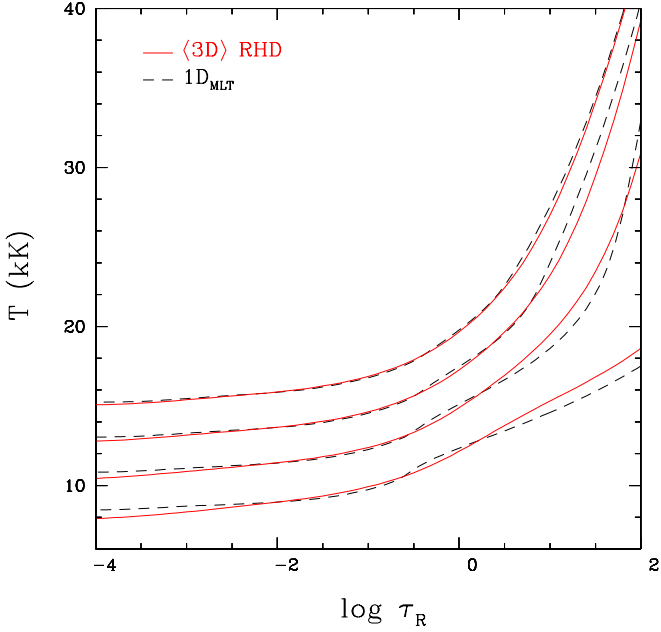}
\caption {Temperatures versus Rosseland optical depth for $\langle$3D$\rangle$ (solid lines) and 1D$_{\rm MLT}$ (dashed lines)
  non-gray simulations at  $T_{\rm eff}=$ 11,300, 11,975, 12,390 and 12,800 K (shifted
  on the vertical axis by 0, 2, 4 and 6 kK, respectively, for
  clarity).
  \label{fg:f3}}
\end{center}
\end{figure}

In order to make our comparison strictly differential, we also employ
1D models in which the microphysics and radiation transfer numerical
schemes (e.g., the opacity binning) are identical to those included
in the 3D models. We computed these hydrostatic 1D models with the LHD
code \citep{caffau07}, which treats convection within the MLT
framework using our most recent ML2/$\alpha=0.8$ parameterization (see
TB10 for details). In Fig. \ref{fg:f3}, we present these models
hereafter named 1D$_{\rm MLT}$ (the gray models are omitted for
clarity). In addition, we computed mean 1D structures that correspond
to the $\langle T^4\rangle ^{1/4}$ and $\langle P \rangle$ spatial
average of the 3D models over surfaces of constant Rosseland optical
depth. We followed this procedure with temporal averages over 12
selected snapshots. These averages, called $\langle$3D$\rangle$ models
and shown in Fig. \ref{fg:f3}, have had their horizontal temperature and
pressure fluctuations removed, and therefore, a comparison of 3D and
$\langle$3D$\rangle$ models highlights the effect of these
fluctuations.

\begin{figure}[!h]
\begin{center}
\includegraphics[bb=58 183 672 648, width=4.5in]{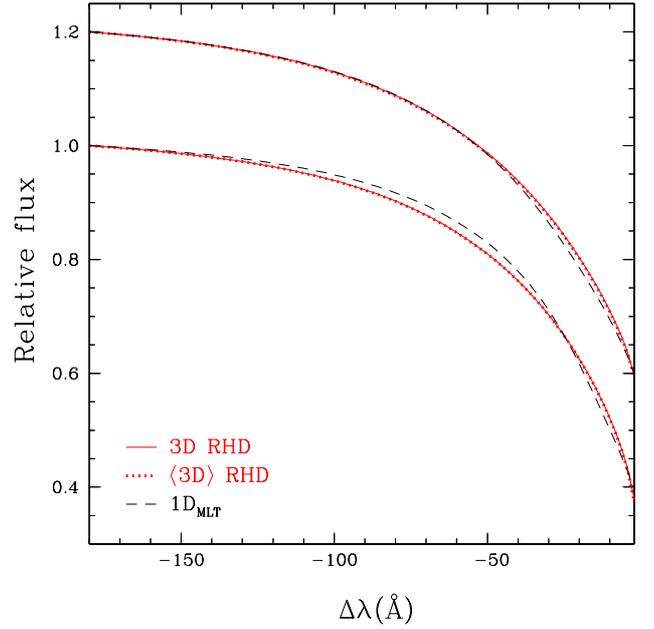}
\caption {Comparison of H$\beta$ line profiles calculated with
  Linfor3D for the 3D (solid lines, averaged over 12 selected
  snapshots) and 1D$_{\rm MLT}$ (dashed lines) models
  in the case of two non-gray simulations at $T_{\rm eff}$ of 11,300 K
  and 12,390 K (the latter shifted by 0.2 units of flux for
  clarity). The dotted lines (indistinguishable from the 3D
  spectra in the figure) also show the $\langle$3D$\rangle$ profiles
  resulting from spatial and temporal averages of the 3D structures.
  \label{fg:f4}}
\end{center}
\end{figure}

At first, we rely on the Linfor3D three-dimensional spectral synthesis
code\footnote{www.aip.de/$\sim$mst/linfor3D$\_$main.html} to compute
H$\beta$ line profiles from our DA model
atmospheres. Figure~\ref{fg:f4} presents the 3D, $\langle$3D$\rangle$,
and 1D$_{\rm MLT}$ profiles at $T_{\rm eff}=11,300$~K and 12,390~K. It
is already obvious that the 3D spectra are indistinguishable from the
spectra computed from the corresponding averaged $\langle$3D$\rangle$
models, an effect also observed by \citet{steffen95} in their 2D
simulations. This is a non-trivial result since significant temperature
and pressure inhomogeneities are present in the hydrodynamical
simulations (see Fig.~\ref{fg:f2}). In the case of main-sequence
stars, for instance, differences between 3D and $\langle$3D$\rangle$
spectra can be important \citep{ludwig09}. However, in the case of DA
white dwarfs, an optical spectrum can be safely computed using a
single averaged atmospheric structure. In this context, one can
therefore neglect the time-consuming task of 3D spectral synthesis.

On the other hand, Fig.~\ref{fg:f4} also demonstrates that the
differences between the $\langle$3D$\rangle$ and 1D$_{\rm MLT}$ models
are significant. This suggests that most of the effects from the RHD
treatment of convection originate from changes in the mean vertical
atmospheric structure. An examination of Fig.~\ref{fg:f3} indeed
reveals that the $\langle$3D$\rangle$ -- hereafter equivalent to 3D --
and 1D$_{\rm MLT}$ structures have a significantly different shape in
the photosphere. For the remainder of this study, our goal is to
quantify the differences between these models.

We chose to compute optical synthetic spectra (3500 \AA~$< \lambda <$
5500 \AA ) for all our 3D and 1D$_{\rm MLT}$ structures calculated in
this work using the standard spectral synthesis code for DA white
dwarfs from the Montreal group (\citealt{TB09}, TB10). The rationale
behind this choice is that, as described above, one can neglect the 3D
spectral synthesis, and moreover, our code relies on an equation of
state and opacity sources that are more appropriate for DA white
dwarfs (e.g., with a proper account of line blending, including
non-ideal gas effects). We verified that the structures calculated in
this work from the LHD and CO$^5$BOLD codes (with the microphysics
dating from \citealt{ludwig94}) are comparable to those computed with
the Montreal code. This implies that the differences in the
microphysics are small, after all, and that our procedure is adequate
for a differential analysis.

\section{Application to the high-{\boldmath $\log g$} problem}

\begin{figure}[!h] \begin{center} \includegraphics[bb=18 160 592
668,width=3.5in]{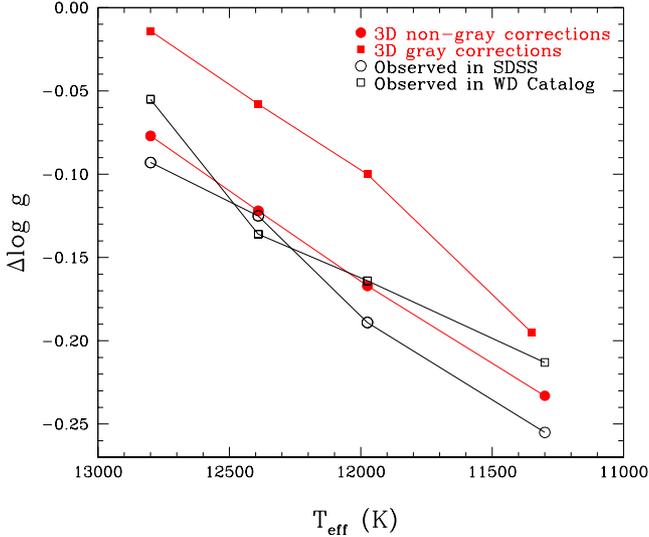} \caption {Non-gray (filled circles) and gray
(filled squares) 3D $\log g$ corrections. In comparison, we show the
shifts in surface gravity required to obtain a stable $\log g$
distribution as a function of $T_{\rm eff}$ using the samples of the
SDSS (open circles, also displayed in Fig.~\ref{fg:f1}) and Gianninas
et al.~(2009; open squares). The points are connected for
clarity. \label{fg:f5}} \end{center} \end{figure}

To quantify the effects of the 3D white dwarf models, we compute in
this section the 3D atmospheric parameter corrections with respect to
the reference 1D$_{\rm MLT}$ models. These are defined as the
differences in $T_{\rm eff}$ and $\log g$ when we fit the normalized
line profiles of both the 1D$_{\rm MLT}$ and 3D spectra with a
reference grid\footnote{We neglect the center of the lines ($\vert
  \Delta \lambda \vert < 3$ \AA), which are formed at low values of
  $\tau_{\rm R}$ where the number of frequencies taken into account in
  the structure calculations can have a strong impact on the
  predictions.}. More specifically, the corrections are defined as
1D$_{\rm MLT}-$3D, since we are interested in how much the atmospheric
parameters of real stars would change when using 3D model spectra. Our
approach is similar to that of \citet{ludwig09} who studied these
corrections for late-type main sequence stars. We use as a reference
grid the spectra computed in TB10\footnote{We also calculated a
  similar reference grid from gray structures to fit our gray
  models.}. Obviously, the atmospheric parameter corrections will not
be {\it exact} since the independent reference grid does not exactly
reproduce the spectra computed in this analysis. Nevertheless, the
good quality of our fits achieved in all cases implies that our
approach should be valid to within a few percents. The 3D $\log g$
corrections are presented in Fig. \ref{fg:f5}. In the case of the
3D corrections in effective temperature, we find that they are all
relatively small ($\sim200$~K at $T_{\rm eff}\sim12,000$~K). They are
certainly of the same order as the accuracy of the spectroscopic
method, and therefore we do not discuss them further. We note that the
3D corrections should approach zero for $T_{\rm eff} \gtrsim 13,000$
K, since the convection zone is rapidly shrinking, and the 1D$_{\rm
  MLT}$ and 3D spectra are otherwise identical.

The non-gray 3D $\log g$ corrections are the key values to assess the
high-$\log g$ problem. A preliminary observation reveals that these
corrections are negative for all spectra (i.e., the 3D models predict
lower surface gravities), which is in the appropriate direction to
correct the problem. The mean amplitude of the corrections is
significant ($\sim$ $-$0.15 dex), which may not be surprising since
the complicated flow patterns are not expected to be fully represented
by a unique free parameter for all atmospheric depths, and for all
white dwarfs. Also given in Fig.~\ref{fg:f5} are the {\it observed}
shifts in $\log g$ derived from the SDSS white dwarf sample displayed
in Fig.~\ref{fg:f1}. These shifts correspond to the corrections
required, in a bin of 1000 K around the model temperature, to match
the median mass value obtained from hot DA stars, represented by the
dashed line in Fig.~\ref{fg:f1}. We also computed similar shifts from
the sample of \citet{gianninas}, drawn from the Villanova White Dwarf
Catalog, to account for observational uncertainties. We can see that
the 3D corrections have {\it almost exactly the right slope and
amplitude} to solve the high-$\log g$ problem. This result clearly
indicates that the surface gravity distribution of cool convective DA
stars ($T_{\rm eff} \gtrsim 11,000$ K) would be much more stable, as a
function of $T_{\rm eff}$, if 3D model atmospheres were used instead
of the 1D standard models.

We also verified that the non-gray $\log g$ corrections are fairly
independent of the $T_{\rm eff}$ corrections by redoing our fits at
fixed temperatures ($T_{\rm eff}=T_{\rm 3D}$). The corrections we
obtain with this method are within $\sim20$\% of those found in
Fig. \ref{fg:f5}. Finally, the fact that we have a set of 3D gray
models allows us to evaluate the impact on our results of the opacity
binning procedure in the non-gray models. The shifts in $\log g$ that
we obtain with 1 opacity bin (gray models) are smaller than those
with 7 bins (non-gray models), but still correct to a large degree the
high-$\log g$ problem. This shows that further improvements in the
binning procedure of the RHD models are not expected to change our
differential analysis significantly.

\section{Conclusion}

We conducted a differential analysis of synthetic spectra for cool
convective DA white dwarfs computed from model atmospheres with a
detailed 3D radiation-hydrodynamical treatment of convective motions,
and with the standard 1D mixing-length theory. We introduced the
concept of 3D atmospheric parameter corrections, which are defined as
the difference in $T_{\rm eff}$ and $\log g$ between 1D$_{\rm MLT}$ and 3D
spectra when fitted with an independent grid. These corrections
illustrate by how much the atmospheric parameters of DA stars would
change if we were to use a grid of 3D spectra to fit the
observations. We find that the 3D surface gravity corrections have the
correct amplitude ($\sim$ $0.15$ dex) to solve the high-$\log g$
problem for DA white dwarfs in the range 11,300 K $< T_{\rm eff} <$ 12,800
K. This indubitably suggests that a weakness with the standard MLT
theory is creating the high $\log g$ values reported in recent
spectroscopic analyses (e.g., TB10).

Our next goal will be to review the microphysics and the opacity
binning procedure in the CO$^5$BOLD code to connect the absolute
properties of these hydrodynamical models with those of standard white
dwarf models. We will then be able to compute improved 3D model
spectra that can actually be compared with observations to determine
the atmospheric parameters. We also plan to look at convective white
dwarfs cooler than those studied in this work, where the MLT theory
predicts an increasingly adiabatic convection, which is less sensitive
to the mixing-length parameter. It remains to be seen how the 3D
$\log g$ corrections would behave in this regime.

\begin{acknowledgements}
This work was supported in part by the NSERC Canada and by the Fund
FQRNT (Qu\'ebec). B.F.\ acknowledges financial support from the {\sl
Agence Nationale de la Recherche} (ANR), and the {\sl ``Programme
Nationale de Physique Stellaire''} (PNPS) of CNRS/INSU, France.
\end{acknowledgements}

\bibliographystyle{aa} 
\bibliography{ms} 

\end{document}